\documentclass{article}
\usepackage{graphicx,psfig,epsfig}
\def\title{\begin{center}\Large\bf}
\def\author(s){\vspace{0.5cm}\large\rm}
\def\text{\end{center}}

\begin{document}

\title
Smaller alignment index (SALI): Determining the ordered or chaotic nature of
orbits in conservative dynamical systems

\author(s)
Ch. Skokos$^{a,b}$, Ch. Antonopoulos$^a$, T. C. Bountis$^a$ \& M. N. Vrahatis$^c$
\footnote{E-mails: hskokos@cc.uoa.gr (Ch.~S.), antonop@math.upatras.gr (Ch.~A.),
bountis@math.upatras.gr (T.~C.~B.), vrahatis@math.upatras.gr (M.~N.~V.)}

\vspace{0.3cm}

$^a${\it Department of Mathematics, Division of Applied Analysis
and
Center for Research and Applications of Nonlinear Systems (CRANS),\\
University of Patras, GR-26500 Patras, Greece}\\
$^b${\it Research Center for Astronomy, Academy of Athens,
14 Anagnostopoulou str.,  GR-10673 Athens, Greece}\\
$^c${\it Department of Mathematics
and University of Patras Artificial Intelligence
 Research Center (UPAIRC),\\
University of Patras, GR-26110 Patras, Greece}\\

\text

\vspace{0.3cm}

\large

\section*{Abstract}
We apply the smaller alignment index (SALI) method for
distinguishing between ordered and chaotic motion in some simple
conservative dynamical systems. In particular we compute the SALI
for ordered and chaotic orbits in a 2D and a 4D symplectic map, as
well as a two--degree of freedom Hamiltonian system due to 
H\'{e}non \& Heiles. In all cases, the SALI determines correctly 
the nature of
 the tested orbit, faster than the 
method of the computation of the maximal Lyapunov characteristic
number. The computation of the SALI for a sample of initial
conditions allows us to clearly distinguish between regions in 
phase space where ordered or chaotic motion occurs.

\section{Introduction}
One of the most important approaches for understanding the
behavior of a dynamical system is based on the knowledge of the chaotic 
vs.~ordered nature of its orbits. For Hamiltonian systems with two
degrees of freedom (or equivalently for 2D symplectic maps), the
inspection  of the consequents of an orbit on a Poincar\'{e}
surface of  section (PSS), can give us reliable information
for the dynamics of individual orbits. On the other hand, the distinction
between ordered and chaotic motion becomes particularly difficult in
systems with many degrees of freedom, where phase space 
visualization is no longer easily accessible.

A quantitative method for distinguishing between order and chaos 
that has been extensively used (also for multidimensional systems),
is the computation of the maximal Lyapunov characteristic number
(LCN) \cite{Benetal, Fr84}. The LCN is the limit of the finite
time Lyapunov characteristic number:
\begin{equation}
L_t=\frac{1}{t}\, \ln  \frac{|\xi_t|}{|\xi_0|}
\label{eq:ftLCN}
\end{equation}
(where $\xi_0$ and $\xi_t$ are the distances between two points of
two nearby orbits at times $t=0$ and $t$), when $t$ tends to
infinity. In other words, LCN measures the average exponential
deviation  of two nearby orbits, so if LCN=0 the tested orbit is
ordered and if \mbox{LCN $>  0$} it is chaotic. In maps, $t$ is a
discrete variable i.e. the number $N$ of iterations of the map, so the
finite time Lyapunov characteristic number can be denoted as
$L_N$. An advantage of this method is that it can be  applied to
systems of any number of degrees of freedom. The main
disadvantage of LCN is that the time needed for $L_t$ to converge
to its limit  can be extremely high and in some cases even
unrealistic for the  systems under study.

A fast, efficient and easy to compute criterion to check if
orbits of multidimensional  maps are chaotic or not
has been introduced in \cite{Sk01}: It concerns the computation 
of the smaller
alignment index (SALI). Recently  this method has been
successfully applied to a two--degree of freedom Hamiltonian flow
\cite{Sketal}. In the present communication we first recall the
definition of the SALI and show its effectiveness by applying it
to a 2D and a 4D symplectic map, comparing it also to the
computation of LCN.  We then use the SALI to study the
dynamics  of the H\'{e}non--Heiles Hamiltonian system;
and illustrate its ability to distinguish between regions of 
the phase space where ordered and chaotic
motion occurs clearly and faster than LCN.

\section{Definition of the smaller alignment index (SALI)}

Let us consider the $2n$--dimensional phase space of a
conservative dynamical system, described by a symplectic map
$\mbox{\bf T}$ or a Hamiltonian system defined by the $n$ degrees
of freedom Hamiltonian function $H$. The time evolution of an
orbit with initial condition $P(0)=(x_1(0),\; x_2(0),\dots,\;
x_{2n}(0))$ is defined by the repeated applications of the map
$\mbox{\bf T}$ or by the solution of Hamilton's equations of motion.

In order to find the LCN one has to compute the limit of the
finite time Lyapunov characteristic number $L_t$ (\ref{eq:ftLCN})
for an initially infinitesimal deviation vector
$\xi(0)=(dx_1(0),\; dx_2(0),\dots,\; dx_{2n}(0))$, as the time $t$
or the number of iterations $N$ tend to infinity, for Hamiltonian
flows and maps respectively. The time evolution of the deviation
vector is given by the equations of the tangent map
\begin{equation}
 \xi(N+1)  =  \left( \frac{\partial \mbox{\bf T}}{\partial P(N)} \right) \xi(N)    ,
\label{eq:Tmap}
\end{equation}
for maps, and by the variational equations 
\begin{equation}
{\bf J} \cdot \dot{\xi}'={\bf DH} \cdot \xi'  , \label{eq:var}
\end{equation}
for flows, where ($'$) denotes the transpose matrix and matrices ${\bf J}$
and ${\bf DH}$ are defined by
\begin{equation}
{\bf J}=\left(
\begin{array}{ccc}
{\bf 0_n} & {\bf -I_n}  \\
{\bf I_n} & {\bf 0_n}
\end{array}
\right ) \; \;, \;\;
DH_{ij}=\frac{\partial^2 H}{\partial x_i \,
\partial x_j}\;\; \mbox{with} \; \; i,j=1,2,...,2n  ,
\end{equation}
 ${\bf I_n}$ being the $n \times n$ identity matrix and ${\bf
0_n}$
 the $n \times n$ matrix with all its elements equal to zero.

In order to define the SALI for the orbit with initial conditions
$P(0)$ we follow the time evolution of two  initial
deviation vectors $\xi_1(0)$ and $\xi_2(0)$. At every time step we
normalize each vector  to $1$ and define the parallel alignment index
\begin{equation}
d_-(t) \equiv   \|\xi_1(t) -\xi_2(t) \|
\label{eq:PALI}
\end{equation}
and the antiparallel alignment index 
\begin{equation}
d_+(t) \equiv   \|\xi_1(t) +\xi_2(t) \|,
\label{eq:APALI}
\end{equation}
following \cite{Sk01} ($\| \cdot \|$ denotes the Euclidean norm of a vector). 
The smaller alignment index SALI is given by
\begin{equation}
\mbox{SALI} = \min (d_-(t),d_+(t)).
\label{eq:SALI}
\end{equation}
From the above definitions we see that when the two vectors
$\xi_1(t)$, $\xi_2(t)$ tend to coincide we get
\begin{displaymath}
d_-(t) \rightarrow 0, \; d_+(t) \rightarrow 2, \; \mbox{SALI}  \rightarrow 0,
\label{eq:coin}
\end{displaymath}
while, when they tend to become opposite we get
\begin{displaymath}
d_-(t) \rightarrow 2, \; d_+(t) \rightarrow 0, \; \mbox{SALI}  \rightarrow 0.
\label{eq:oppo}
\end{displaymath}
So, it is evident that SALI is a quantity that clearly informs us
if the two deviation vectors tend to have the same direction by
coinciding or becoming opposite. The reason why this information
is useful for understanding if an orbit is chaotic or not is that,
for systems of $2n$--dimensional phase space with $n\geq 2$, the
two vectors tend to coincide or become opposite for chaotic orbits
\cite{Vogetal}, i.e. the SALI tends to zero. This 
is due to the fact that the direction of the two deviation vectors
tends to coincide with the direction of the most unstable
nearby manifold. On the other hand, if the tested orbit is ordered
it lies on a torus and the two deviation vectors eventually become
tangent to the torus, but in general converge to different directions.
In that case the SALI does not tend to zero, but its values
fluctuate around a positive value.

Although the SALI method is perfectly suited for multidimensional
systems, it can also be applied to 2D maps. For 2D maps, whose
phase space is 2--dimensional, the SALI tends to zero both for
ordered and chaotic orbits, but with completely different time
rates which allows us to clearly distinguish between the two
cases. This behavior is due to the fact that for ordered orbits
the two vectors, as has already mentioned, become tangent to the
torus, which now is simply an invariant  curve. So, the only
possibilities for the two vectors are to become identical or
opposite which means that the SALI tends to zero.

\section{Applications of the SALI}

\subsection{Symplectic maps}

Following \cite{Sk01} we compute the SALI in some simple cases of
ordered and chaotic orbits in symplectic maps with two and four
dimensions. In particular we  use the 2D map:
\begin{equation}
\begin{array}{lll}
 x'_1  &=&  x_1 + x_2   \\
 x'_2  &=&  x_2 - \nu \sin (x_1 + x_2)
\end{array}
\pmod{2 \pi} ,
\label{eq:2Dmap}
\end{equation}
and the 4D map:
\begin{equation}
\begin{array}{l}
 x'_1  =  x_1 + x_2   \\
 x'_2  =  x_2 - \nu \sin (x_1 + x_2) - \mu [1- \cos (x_1+x_2+x_3+x_4)]    \\
 x'_3  =  x_3 + x_4   \\
 x'_4  =  x_4 - \kappa \sin (x_3 + x_4) - \mu [1- \cos (x_1+x_2+x_3+x_4)]
\end{array}
\pmod{2 \pi},
\label{eq:4Dmap}
\end{equation}
which is composed of two 2D maps of the form (\ref{eq:2Dmap}),
with parameters $\nu$ and $\kappa$, coupled with a term of order
$\mu$. All variables are given (mod $2\pi$), so \mbox{$x_i \in
[-\pi,\pi),$} for \mbox{$i=1,2,3,4$}. The map (\ref{eq:4Dmap}) is
a variant of the 4D map studied by Froeschl\'{e} \cite{Fr72}. Some
dynamical structures on the phase space of this map were examined
in detail in \cite{Sketal77} for small values of the coupling
parameter $\mu$.

\begin{figure}
\centerline{\epsfxsize=14.0cm \epsfbox{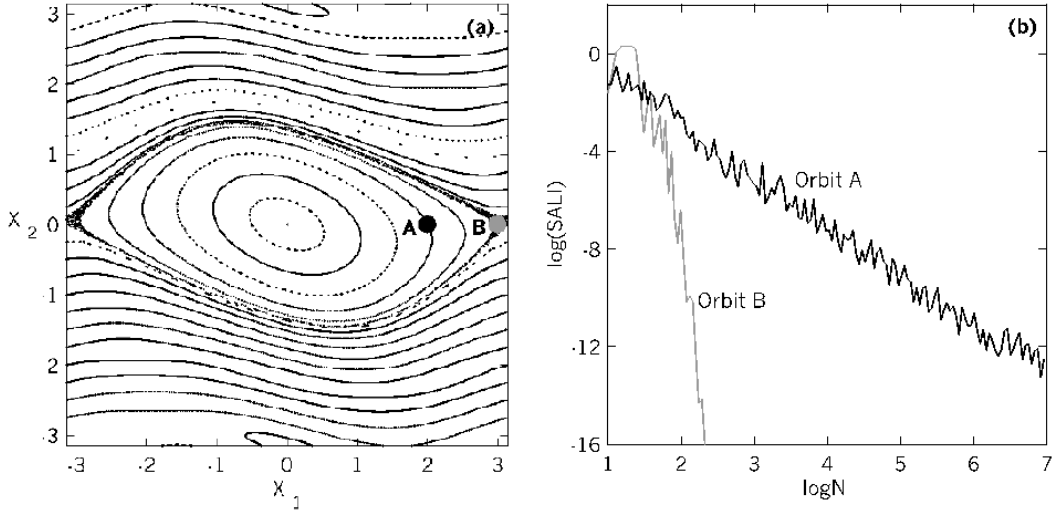}}
\caption{ (a)
Phase plot of the 2D map (\ref{eq:2Dmap}) for $\nu=0.5$. The
initial conditions of the ordered orbit A ($x_1=2,\; x_2=0$)  and
the chaotic orbit B ($x_1=3,\; x_2=0$) are marked by black and
light-gray filled circles respectively. (b) The evolution of the
SALI, with respect to the  number $N$ of iterations of the 2D map
(\ref{eq:2Dmap}) for orbit A (solid line) and for orbit B (dashed
line). }
\label{f:2D}
\end{figure}

In the case of the 2D map (\ref{eq:2Dmap}) we consider the ordered
orbit A with initial condition $x_1=2$, $x_2=0$ and the chaotic
orbit B with initial condition $x_1=3$, $x_2=0$ for $\nu=0.5$. The
phase plot of map (\ref{eq:2Dmap}) can be seen in figure
\ref{f:2D}(a), where the initial conditions of orbits A and B are
marked by black and light-gray circles, respectively. The ordered
behavior of orbit A and the chaotic nature of orbit B are evident
from the distribution of their consequents on the 2D phase space.
In particular the successive consequents of orbit A lie on a
smooth invariant curve, while the successive consequents of orbit
B are scattered in the small chaotic region that surrounds the
main stability island around $x_1=x_2=0$. The different nature of
the two orbits is revealed also by the behavior of the SALI. The
initial deviation vectors used for the computation of the SALI are
$\xi_1(0)=(1,0)$ and $\xi_2(0)=(0,1)$ for both orbits. These
vectors eventually coincide in both cases, but at completely
different time rates. This is evident in  figure \ref{f:2D}(b),
where the SALI is plotted as a function of the number N of
iterations for the ordered orbit A (black line) and the chaotic
orbit B (gray line) in log-log scale. For the ordered orbit A the
SALI decreases as N increases, following a power law and becomes
$\mbox{SALI}\approx 10^{-13}$ after $10^7$ iterations. On the
other hand, the SALI of the chaotic orbit B decreases abruptly,
reaching the limit of accuracy of the computer ($10^{-16}$) after
only about 200 iterations. After that time, the two vectors become
identical to computer accuracy. 
So, it becomes evident that the SALI can distinguish
between ordered and chaotic motion even in a 2D map, since it tends to
zero following completely different time rates.

\begin{figure}
\centerline{%
\begin{tabular}{c@{\hspace{2pc}}c}
\includegraphics[width=6.7cm]{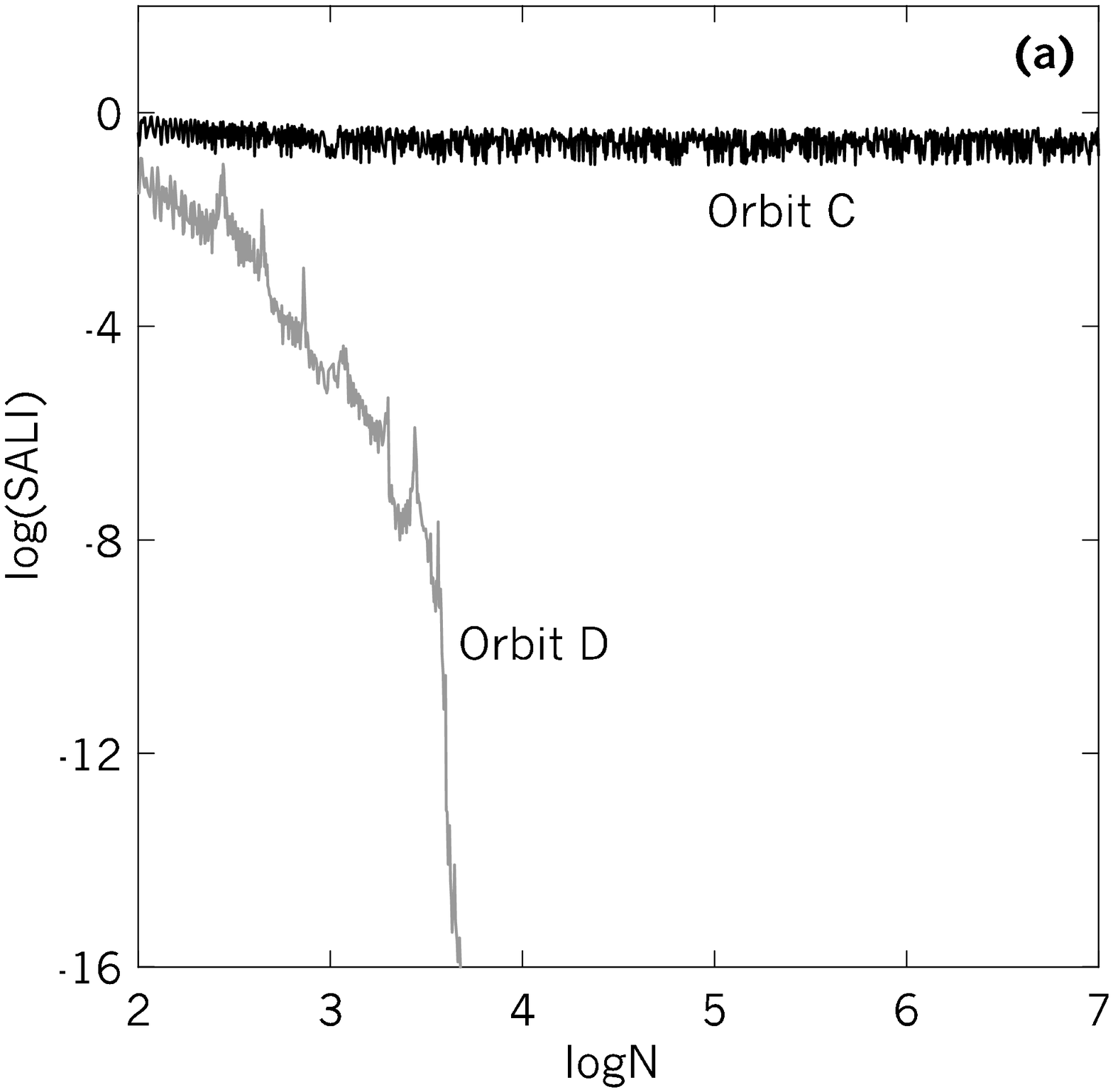} &
\includegraphics[width=6.7cm]{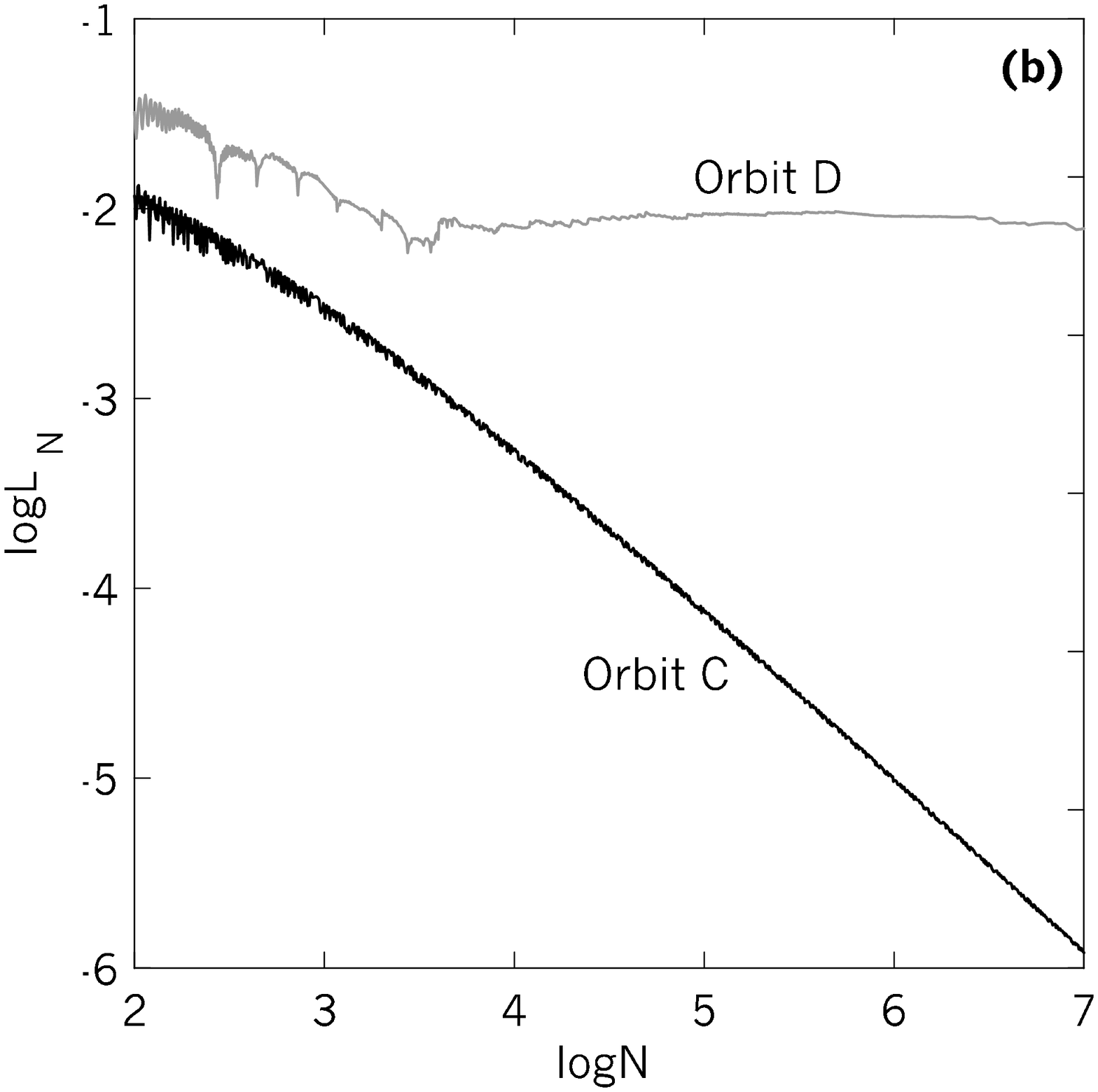} \\
\end{tabular}}
\caption{ The evolution in log-log scale of (a)  the  SALI and (b)
the finite time Lyapunov characteristic number $L_N$, with respect
to the  number $N$ of iterations of the  4D map  (\ref{eq:4Dmap})
with $\nu=0.5$, $\kappa=0.1$,  $\mu=10^{-3}$, for the ordered
orbit C with initial condition $x_1=0.5$,  $x_2=0$, $x_3=0.5$,
$x_4=0$ (black line) and for the chaotic orbit D with  initial
condition $x_1=3$, $x_2=0$, $x_3=0.5$, $x_4=0$ (gray line). }
\label{f:4D}
\end{figure}

In the case of the 4D map (\ref{eq:4Dmap}) for $\nu=0.5$,
$\kappa=0.1$ and $\mu=10^{-3}$ we consider the ordered orbit C
with initial condition $x_1=0.5$, $x_2=0$, $x_3=0.5$, $x_4=0$ and
the chaotic orbit D with initial condition $x_1=3$, $x_2=0$,
$x_3=0.5$, $x_4=0$. The initial deviation vectors used for the
computation of the SALI are (1,1,1,1) and (1,0,0,0), for both
orbits. As we see in figure \ref{f:4D}(a) the SALI of the ordered
orbit C remains almost constant (black line), fluctuating around
$\mbox{SALI} \approx 0.28$. On the other hand, the SALI of the
chaotic orbit D decreases abruptly, reaching the limit of accuracy
of the computer ($10^{-16}$) after about $4.7\times 10^3$
iterations (gray line). After that time, the coordinates of the two
vectors are represented by opposite numbers in the computer (since
the SALI actually coincides with $d_+$ in this case), and any
further computation of their evolution is not necessary. 

So, in 4D
maps the SALI tends to zero for chaotic orbits, while it tends to
a positive value for ordered orbits. Thus, the different behavior
of SALI clearly distinguishes between ordered and chaotic motion.
Another advantage of using the SALI is that we can be sure about
the nature of the tested orbit faster than using LCN. This becomes
evident by looking at the evolution of the finite time Lyapunov
characteristic number $L_N$ (\ref{eq:ftLCN}) for orbits C and D in
figure \ref{f:4D}(b). As expected, for the ordered orbit C, $L_N$ 
 decreases as the number of iterations $N$ increases,
following a power law, reaching the value $L_N \approx 1.6\times
10^{-6}$ after $10^7$ iterations. On the other hand, $L_N$ of the
chaotic orbit D, after some fluctuations, seems to stabilize near a
constant non--zero value  $L_N \approx 5 \times
10^{-2}$ after $10^7$ iterations. By comparing panels (a) and (b)
of figure \ref{f:4D} we see that, after about $4.7\times 10^3$
iterations we can be sure that orbit D is chaotic using the SALI,
since it has become equal to $10^{-16}$ and the two deviation
vectors practically coincide, while we cannot stop the computation
of $L_N$ at that time, since it is not yet evident whether the LCN 
for orbit D will ultimately be zero or not.

\subsection{The H\'{e}non--Heiles Hamiltonian system}

\begin{figure}
\centerline{\epsfxsize=10.0cm \epsfbox{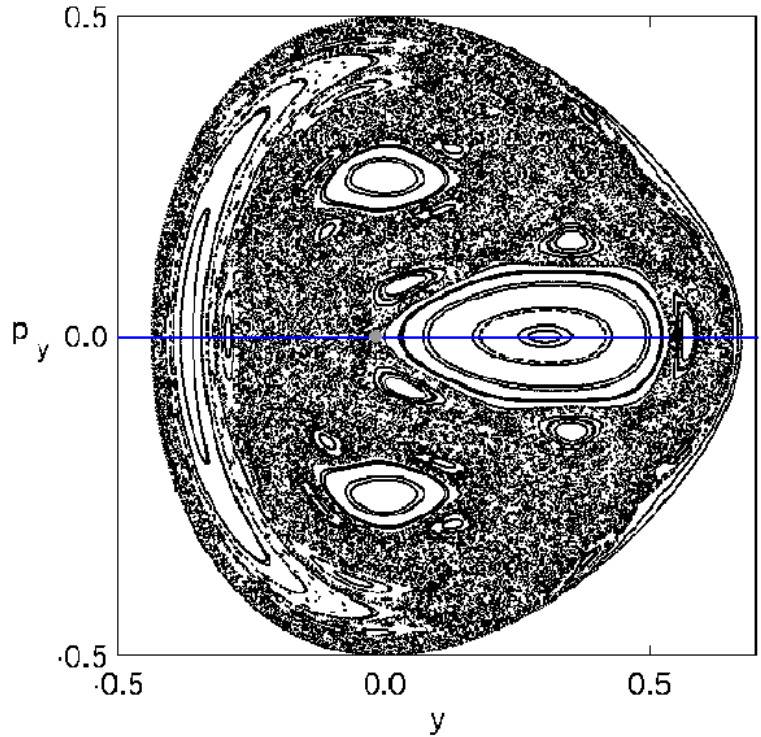}}
\caption{ The
Poincar\'{e} surface of section for $x=0$ of the two degrees of
freedom H\'{e}non--Heiles Hamiltonian (\ref{eq:2DHam}) for H=1/8.
The projection on the PSS of the initial conditions of the ordered
orbit E ($x=0,\; y=0.55,\; p_x\simeq0.2417 ,\; p_y=0$) and the
chaotic orbit F ($x=0,\; y=-0.016,\; p_x\simeq0.49974 ,\; p_y=0$)
are marked by black and light-gray filled circles respectively.
The axis $p_y=0$ is also plotted. }
\label{f:PSS}
\end{figure}

In order to illustrate the effectiveness of the SALI in
determining  chaotic vs.~ordered  orbits in Hamiltonian
flows, we consider the two degrees of freedom H\'{e}non--Heiles
Hamiltonian \cite{HenHe}
\begin{equation}
H(x, y, p_x, p_y) = \frac{1}{2} (p_x^2+p_y^2) + \frac{1}{2} (x^2+y^2)
+ x^2 y - \frac{1}{3} y^3,
\label{eq:2DHam}
\end{equation}
where $x$, $y$ are the generalized coordinates and $p_x$, $p_y$
the conjugate momenta. In particular, we consider the case of fixed energy
$H=1/8$, for which the system exhibits a rich dynamical structure.
As it can be seen on the Poincar\'{e} surface of section (PSS) for
$x=0$ in figure \ref{f:PSS} there exist islands of stability,
where ordered motion occurs, as well as extensive regions where
chaotic motion takes place.

\begin{figure}
\centerline{%
\begin{tabular}{c@{\hspace{2pc}}c}
\includegraphics[width=6.7cm]{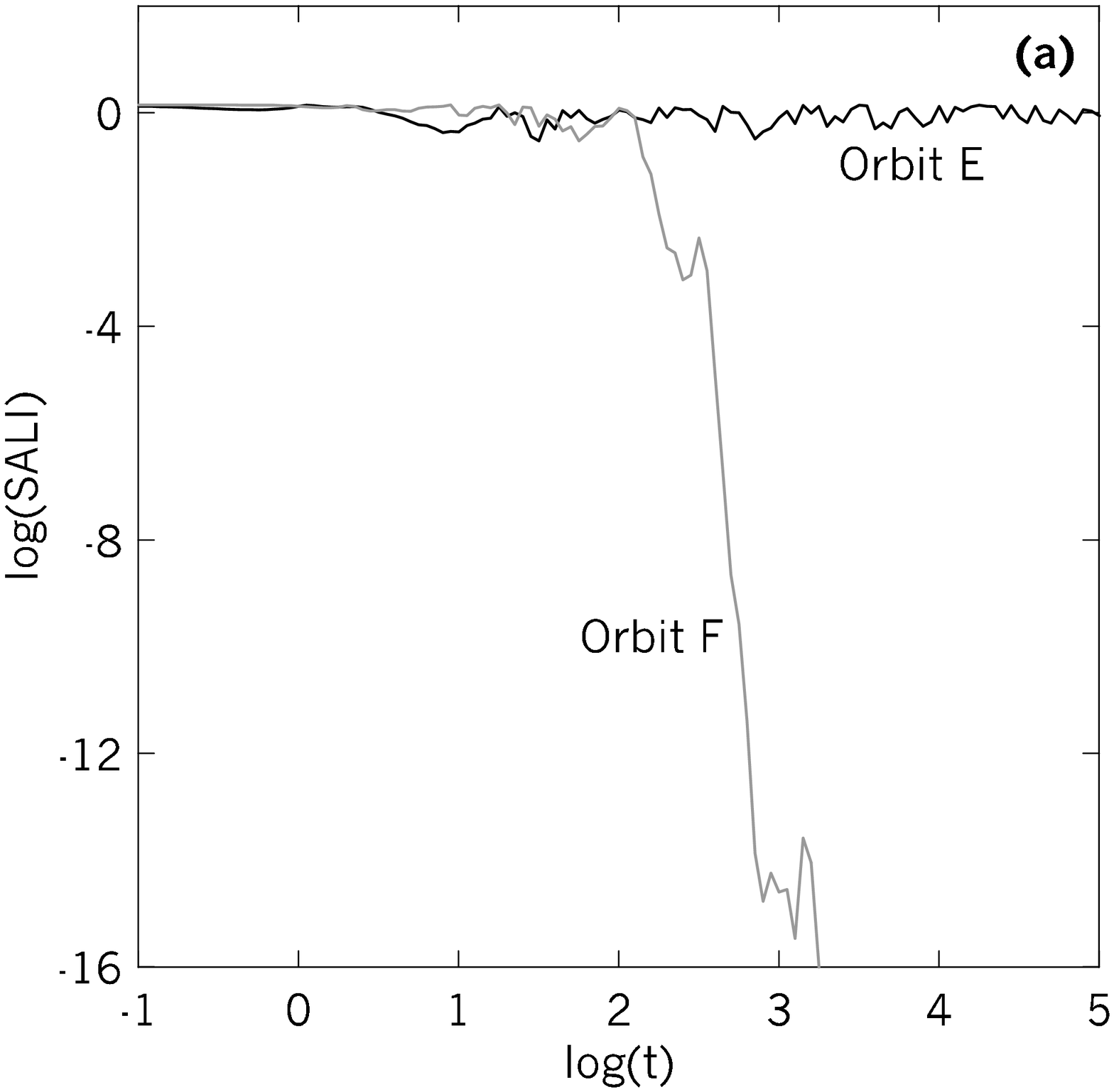} &
\includegraphics[width=6.7cm]{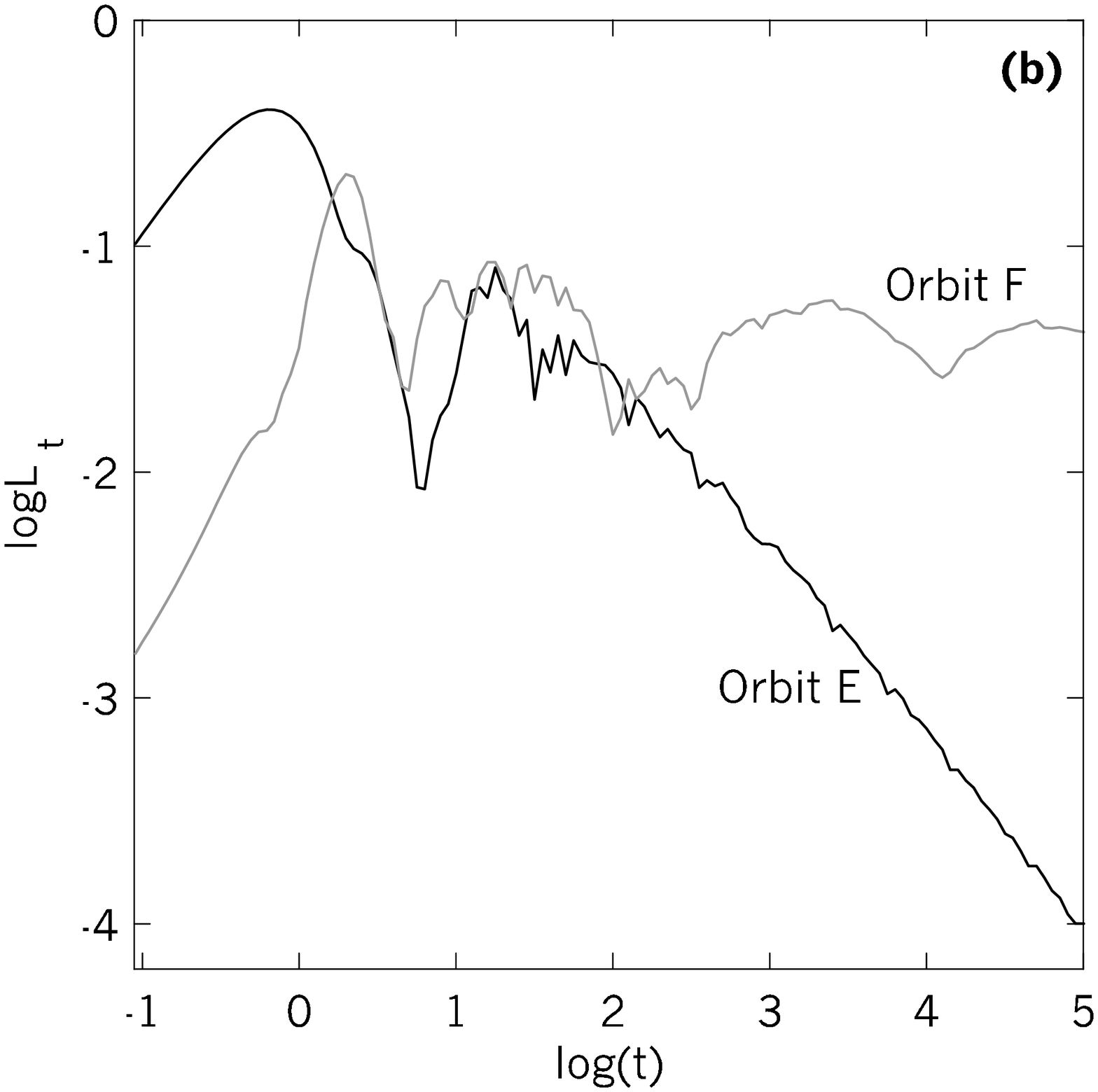} \\
\end{tabular}}
\caption{ The evolution in log-log scale of (a)  the  SALI and (b)
the finite time Lyapunov characteristic number $L_t$ as a function
of time $t$, for the  Hamiltonian  (\ref{eq:2DHam}) with H=1/8,
for the ordered orbit E with initial condition $x=0,\; y=0.55,\;
p_x\simeq0.2417 ,\; p_y=0$ (black line) and for the chaotic orbit
F with initial condition $x=0,\; y=-0.016,\; p_x\simeq0.49974 ,\;
p_y=0$ (gray line). } \label{f:HSALI}
\end{figure}

In order to apply the SALI method, we consider the ordered orbit E
with initial condition $x=0,\; y=0.55,\; p_x\simeq0.2417 ,\;
p_y=0$ and the chaotic orbit F with initial condition $x=0,\;
y=-0.016,\; p_x\simeq0.49974 ,\; p_y=0$. The projection of the
initial conditions of orbits E and F on the PSS are marked by
black and light-gray filled circles respectively in figure
\ref{f:PSS}. The initial deviation vectors $(dx(0),dy(0), dp_x(0),
dp_y(0))$ used for the computation of SALI are $(1,0,0,0)$ and
$(0,0,1,0)$. 

As we see in figure \ref{f:HSALI}(a) the SALI of the
ordered orbit E remains almost constant, fluctuating around
$\mbox{SALI} \approx 1$ (black line), while the SALI of the
chaotic orbit F decreases abruptly reaching the limit of accuracy
of the computer $(10^{-16})$ after about 1,700 time units (gray
line). The behavior of the SALI is similar to the one encountered
in the  4D map (\ref{eq:4Dmap}) (figure \ref{f:4D}(a)),
since the phase space of the Hamiltonian system is 4--dimensional
as in the case of the 4D map. In figure \ref{f:HSALI}(b) we see the time
evolution of finite time Lyapunov characteristic number $L_t$
(\ref{eq:ftLCN}) for orbits E (black line) and F (gray line).
$L_t$ of the ordered orbit E, after an initial transient time
interval where it exhibits large fluctuations, starts to decrease
following a power law, reaching the value $L_t \approx  10^{-4}$
for $t=10^5$. $L_t$ of the chaotic orbit F has larger
fluctuations without tending to zero, although even for $t=10^5$
it does not seem to stabilize around a non-zero value. 

We
underline the fact that in this case, using the SALI we were
absolutely sure that orbit F is chaotic at $t=1,700$, at which
time SALI became practically zero, although at that time the use
of $L_t$ could not give us the same information.

In figure \ref{f:PSS} we see that in a large portion of  phase
space the motion of system (\ref{eq:2DHam}) is chaotic. 
The chaotic region corresponds to the
area filled with scattered points on the PSS, while ordered motion
corresponds to the islands formed by the invariant smooth curves.
Since  the SALI tends to completely different values
for ordered and chaotic orbits, its computation for a sample of
initial conditions can be used to distinguish between regions
where ordered or chaotic motion occurs.

\begin{figure}
\centerline{\epsfxsize=14.0cm \epsfbox{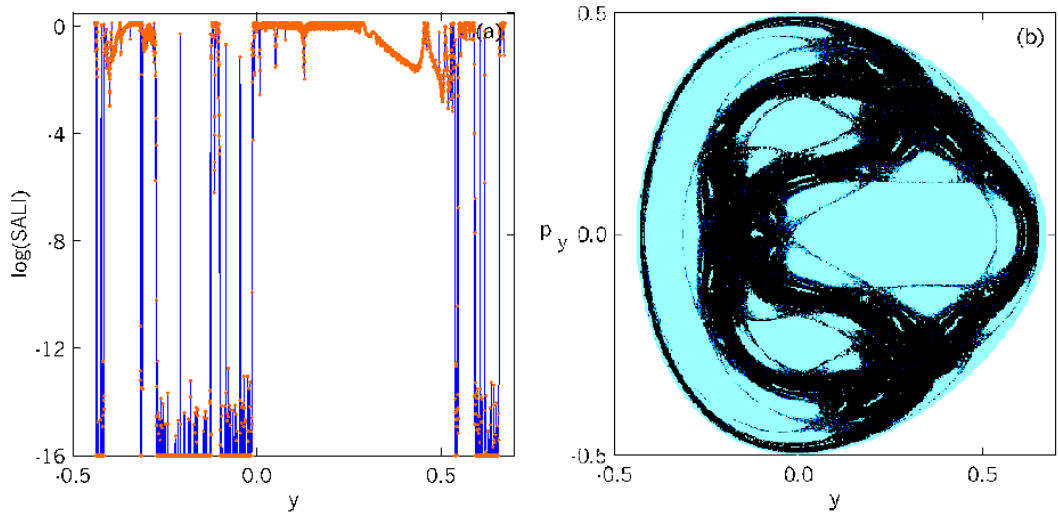}}
\caption{ (a)
The value of SALI for $t=4,000$ for orbits with initial conditions
on the $p_y=0$ axis of the PSS shown in figure \ref{f:PSS}, as a
function of the $y$ variable of the initial condition. The data
are plotted as red points and are connected by blue lines. (b)
Regions of different values of the SALI on the PSS $(y, p_y)$
after 1,000 time steps. Initial conditions that give
$\mbox{SALI}<10^{-12}$ are marked by black points, initial
conditions that give $10^{-12}\leq \mbox{SALI}<10^{-8}$ are marked
by deep blue points, initial conditions that give $10^{-8}\leq
\mbox{SALI}<10^{-4}$ are marked by blue points, while initial
conditions that give $10^{-4}\leq \mbox{SALI}$ are marked by light
blue points. }
\label{f:color}
\end{figure}

As a first example, we consider orbits that lie on the  $p_y=0$
line on the PSS shown in figure  \ref{f:PSS}, having initial
conditions $x=0$, $p_y=0$, while $p_x$ is defined by the
Hamiltonian (\ref{eq:2DHam}). The values of the SALI for all these
orbits, after 4,000 time units, are plotted with red points in
figure \ref{f:color}(a), as function of the $y$ variable of the
initial condition. These points are line connected in order to be
easily visible the changes of the SALI as the initial condition
moves on the $p_y=0$ line. We can clearly see regions of ordered
motion where the SALI has  values larger than $10^{-4}$,
 corresponding to the islands of stability that are crossed by
the line $p_y=0$ in  figure  \ref{f:PSS}. There also exist regions
of chaotic motion where  the SALI has become
smaller than  $10^{-12}$ or has even reached the limit of accuracy
of the computer $(10^{-16})$, in agreement to the regions crossed
by the  $p_y=0$ line where scattered points exist on the PSS.
Although most of the initial conditions give large $(\geq
10^{-4})$ or very small $(<10^{-12})$ values for the SALI, there
also exist  initial conditions that give, after 4,000 time units,
intermediate values for the SALI $(10^{-12}\leq \mbox{SALI} <
10^{-4})$. These  correspond to  sticky orbits existing
near the borders of ordered motion, and more time is needed 
for the SALI to reach
 very small values and reveal their chaotic nature.

By carrying out the above analysis not only on a line on the PSS but for
the whole plane, plotting with different colors initial
points that give values for the SALI in different ranges, we can
get an image of  phase space  regions where ordered and
chaotic motion are clearly distinguished (figure
\ref{f:color}(b)). In figure \ref{f:color}(b) we see that our PSS
is practically divided into  regions where ordered
motion occurs, colored in light blue, which corresponds to
$10^{-4}\leq \mbox{SALI}$,  and  those  colored in black, where chaotic
behavior occurs,   corresponding to
$\mbox{SALI}<10^{-12}$. On the borders between these two regions
we see points that give intermediate values for the SALI colored
in deep blue ($10^{-12}\leq \mbox{SALI}<10^{-8}$) and in blue
($10^{-8}\leq \mbox{SALI}<10^{-4}$), which correspond to sticky
orbits. The resemblance between figure \ref{f:color}(b) and figure
\ref{f:PSS} is obvious. We should also mention that in figure
\ref{f:color}(b), we can see some very thin regions of ordered
motion  corresponding to very small islands of stability that
cannot be seen easily  in figure \ref{f:PSS}. So, it is
evident that starting with any initial condition, the computed
value of the SALI rapidly gives a clear view of chaotic 
vs.~ordered motion even for systems described by ordinary
differential equations, where surface of section plots are already
time consuming for 2 degrees of freedom and practically useless
for systems of higher dimensionality.

\section{Conclusions}

In this paper, we have given some examples of symplectic maps
 and Hamiltonian systems, where the computation of the smaller
alignment index SALI allows us to distinguish in a  cost effective
way between ordered and chaotic orbits.

The computation of the SALI is a fast, efficient and easy to
compute numerical method, perfectly suited for multidimensional
systems, but it can also be applied to 2D maps. In 2D maps,  the
SALI tends to zero both for ordered and chaotic orbits, but
following completely different time rates which allows us to
distinguish between the two cases. In maps of higher
dimensionality (and Hamiltonian systems) the SALI tends to zero
for chaotic orbits, while in general, it tends to a positive
 value for ordered orbits. So, we can easily distinguish
between regular and chaotic orbits. Our approach, in fact, begins
to be truly valuable for Hamiltonian systems of 2 degrees of
freedom (where detailed surface of section plots are
computationally too costly) and promises to become extremely
useful for higher than 2 degree of freedom Hamiltonians and higher
dimensional symplectic maps.

An advantage of using the SALI in Hamiltonian systems or in
multidimensional maps is that usually the chaotic nature of an
orbit can be established beyond any doubt. This happens because
when the orbit under consideration is chaotic, the SALI becomes
 zero, in the sense that it reaches the limit of the
accuracy of the computer. After that time the two deviation
vectors, needed for the computation of the SALI, are identical
(equal or opposite), since their coordinates are represented by
the same or opposite numbers. Thus, they have
exactly the same evolution in time and cannot be separated. This
practically means that we do not need to continue the computation
of the evolution of the two vectors further on.

We should also mention that in all  cases  studied, the use of
the SALI helped us decide if an orbit is ordered or chaotic much
faster than the computation of the finite time Lyapunov
characteristic number $L_t$.

\section*{Acknowledgments}
Ch.~Skokos thanks the LOC of the conference for its financial
support. Ch.~Skokos was also supported by `Karatheodory'
post--doctoral fellowship No 2794 of the University of Patras and
by the Research Committee of the Academy of Athens (program No
200/532). Ch. Antonopoulos was supported by  `Karatheodory'
graduate student fellowship No 2464 of the University of Patras.

\end{document}